\documentstyle[12pt]{article}

\begin{document}
\rightline{CPP-94-18}
\rightline{BNL-60339}
\rightline{ILL-(TH)-94-9}
\bigskip\bigskip
\begin{center}
{\Large \bf Higgs Decay to Top Quarks at Hadron Colliders} \\
\bigskip\bigskip\bigskip\bigskip
{\large\bf D.\ Dicus}\\
\medskip
Center for Particle Physics \\
University of Texas \\
Austin, TX\ \ 78712 \\
\bigskip\bigskip
{\large\bf A.\ Stange} \\
\medskip
Physics Department  \\
Brookhaven National Laboratory \\
Upton, NY\ \ 11973  \\
\bigskip\bigskip
{\large\bf S.\ Willenbrock} \\
\medskip
Physics Department \\
University of Illinois \\
1110 W. Green Street \\
Urbana, IL\ \ 61801 \\
\end{center}
\bigskip\bigskip\bigskip

\begin{abstract}
Higgs bosons which decay principally to top quarks, such as in the minimal
supersymmetric model, produce a peak-dip structure in the $gg\to t\bar t$
invariant-mass spectrum. This structure is potentially observable
at the CERN Large Hadron Collider.
\end{abstract}

\newpage

\addtolength{\baselineskip}{9pt}

\section*{}

\indent\indent In the standard model of the electroweak interaction, the
electroweak symmetry is broken by a scalar $SU(2)_L$ doublet Higgs field which
acquires a vacuum-expectation value.  This mechanism provides masses for the
weak
vector bosons, as well as for the quarks and leptons.  However, the mechanism
which is responsible for the quark and lepton masses need not be as closely
linked to the mechanism which generates the weak-vector-boson masses
as it is in the standard model.  Since fermion masses violate the
$SU(2)_L$ symmetry, whatever provides their masses must break this
symmetry, and contribute to the weak-vector-boson masses.  However, this
contribution may be only a small component of the weak-vector-boson masses,
which may arise dominantly via another mechanism \cite{BC}.

An explicit realization of this scenario is provided by a two-Higgs-doublet
model.  One doublet may acquire a much larger vacuum-expectation value
than the other doublet, and thereby provide the dominant contribution to
the weak-vector-boson masses.  However, this doublet need not couple to
fermions, in which case the fermion masses arise solely from the Higgs
doublet with a small vacuum-expectation value \cite{HKS}.

A two-Higgs-doublet model
produces a spectrum of physical particles which consists of two neutral
Higgs scalars, one neutral pseudoscalar, and a charged Higgs boson
\cite{HKS,HHG}.
The Higgs scalar most closely associated with the doublet with small
vacuum-expectation value couples very weakly
to the weak vector bosons, but couples with enhanced strength to the
fermions (with respect to the standard-model coupling).
The pseudoscalar Higgs particle does not couple to the weak
vector bosons at all, but couples to the fermions with enhanced strength.

In this letter we consider searching for a Higgs
boson, with zero or suppressed coupling to the weak vector bosons, via its
decay to $t\bar t$.  This particle is copiously produced at a hadron
collider through gluon-gluon collisions, via a virtual top-quark loop
\cite{GGMN}, as shown in Fig.~1(a).
However, at a hadron collider there is a large irreducible background
from the QCD production of top quarks, shown in Fig.~1(b).
We will show that the
signal, $gg\rightarrow H\rightarrow t\bar t$, and the background, $gg
\rightarrow t\bar t$, interfere, generically resulting in a peak-dip
structure at the Higgs mass. This phenomenon was first recognized in
Ref.~\cite{GH}. In some cases the dip dominates, such that
the signal for the presence of the Higgs boson is a small {\it deficit} in the
production of $t\bar t$ pairs of invariant mass near the Higgs mass.

Another model which yields Higgs bosons with suppressed coupling to weak
vector bosons is the Higgs sector of the minimal supersymmetric model
\cite{HHG}.
This model necessarily has two Higgs doublets, one which couples to
fermions with $T_{3L}=+1/2$ and one which couples to fermions with
$T_{3L}=-1/2$.  As usual, the pseudoscalar Higgs boson does not couple to
the weak vector bosons.  If this pseudoscalar is relatively heavy
(greater than about 100-150 GeV, depending on the ratio of the
vacuum-expectation values of the two Higgs doublets), the heavier Higgs scalar
(whose mass is then close
to that of the pseudoscalar) couples only weakly to the weak vector bosons.
This is true regardless of the ratio of vacuum-expectation
values of the two Higgs doublets.

For definiteness, we consider the Higgs sector of the minimal
supersymmetric model, with the ratio of the vacuum-expectation values of
the two Higgs doublets close to unity.  The heavier Higgs scalar, $H$,
nearly decouples from the weak vector bosons if its mass is greater than about
150 GeV, and its coupling to top quarks is close to standard-model
strength.  Our qualitative results are generic to the scenario where the
dominant
decay of the Higgs boson is to top quarks. If the ratio of vacuum-expectation
values is much larger than
unity, the coupling of $H$ to the top quark is suppressed and its
coupling to the bottom quark enhanced, such that
the production of the Higgs via a bottom-quark loop, and its decay to
$b\bar b$, become non-negligible.  In the case of a non-supersymmetric scalar
Higgs boson from a two-doublet model, the decay of the Higgs to weak vector
bosons may also be non-negligible.  Both of these scenarios decrease the
branching ratio of $H\to t\bar t$, and hence decrease the size of the effect of
the Higgs boson on top-quark production.

The differential cross section for $gg\rightarrow t\bar t$,
including the squares of the scalar-Higgs amplitude and the
continuum QCD amplitude, as well as the interference of the two amplitudes, is
\footnote{The total cross section may be obtained from this expression
by integrating over $z$, the cosine of the scattering angle between a gluon
and the top quark.  The resulting expression agrees with that of Eq.~7 of
Ref.~\cite{GH}, except the interference term, which is a factor of 2 too
large in that paper. The resulting
cross section, Fig.~2 in that paper, should display a large peak,
as shown, followed by a small dip, rather than an equally-large peak and dip.}
\begin{eqnarray}
\frac{d\sigma}{dz} & = & \frac{\alpha_s^2 G_F^2m^2s^2}{1536 \pi^3}\beta^3
\left|\frac{N(s/m^2)}{s-m_H^2+im_H\Gamma_H(s)}\right|^2 \nonumber \\
& - & \frac{\alpha_s^2 G_Fm^2s}{384\pi\sqrt 2}\beta^3 \left(\frac{1}{p_1\cdot
p_3}+\frac{1}{p_2\cdot p_3}\right)
{\rm Re}\left[\frac{N(s/m^2)}{s-m_H^2+im_H\Gamma_H(s)}\right] \nonumber \\
& + & \frac{d\sigma_{QCD}}{dz}
\end{eqnarray}
where $p_{1,2}$ are the momenta of the incoming gluons, $p_{3,4}$ are
the outgoing top-quark and top-antiquark momenta,
$z$ is the cosine of the scattering angle between an incoming gluon
and the top quark, $m$ is the top-quark mass, and $\beta\equiv
(1-4m^2/s)^{1/2}$
is the velocity of the top quark and top antiquark in the
center-of-momentum frame. The dot product of
the four momenta are
\begin{eqnarray}
p_1\cdot p_3 & = & \frac{s}{4}(1-\beta z) \nonumber \\
p_2\cdot p_3 & = & \frac{s}{4}(1+\beta z)\;.
\end{eqnarray}
The function associated with the virtual top-quark loop is
\begin{equation}
N(s/m^2)=\frac{3}{2} \frac{m^2}{s}\left[4-\left(1-\frac{4m^2}{s}\right)
I(s/m^2)\right]
\end{equation}
where
\begin{equation}
I(s/m^2)=\left[\ln\frac{1+\beta}{1-\beta}-i\pi\right]^2 \quad(s>4m^2)\;.
\end{equation}
The energy-dependent Higgs width is
\begin{equation}
m_H\Gamma_H(s)=\frac{3G_Fm^2s}{4\pi\sqrt 2}\beta^3\;.
\end{equation}
The cross section for the continuum QCD production of $gg\to t\bar t$ is
\cite{QCD}
\begin{equation}
\frac{d\sigma_{QCD}}{dz}=\frac{\pi\alpha_s^2}{12s}\beta
\left(\frac{s^2}{p_1\cdot p_3p_2\cdot p_3}-9\right)
\left[\frac{(p_1\cdot p_3)^2}{s^2}+\frac{(p_2\cdot p_3)^2}{s^2}+\frac{m^2}{s}
-\frac{m^4}{4p_1\cdot p_3p_2\cdot p_3}\right]\;.
\end{equation}

We show in Fig.~2 the cross section for $gg\rightarrow t\bar t$ as a
function of the $t\bar t$ invariant mass, $\sqrt s$, for $m_t=$ 170 GeV
(the approximate central value from precision
electroweak experiments \cite{EW})\footnote{The cross section in the
vicinity of the Higgs mass depends only on the ratio $m_t/m_H$, so the
qualitative results for other top-quark masses can be obtained by scaling the
Higgs mass, modulo the effect of the change in the Higgs width.}
and $m_H=$ 400, 500, 600, 700, and 800 GeV.  For $m_H=$ 400 GeV, the Higgs
boson produces a narrow peak with a width of about 3.3 GeV.
For $m_H=$ 500 and 600 GeV, the presence of the Higgs boson
produces a peak, followed by a dip, near the Higgs mass.  For larger Higgs
masses the peak is absent, and the Higgs reveals itself as a dip in the
$t\bar t$ invariant-mass spectrum.
For $m_H=$ 500 GeV, the {\it total} top-quark cross section differs
little from the cross section with no Higgs present, due to the
cancellation between the peak and the dip.  For larger Higgs masses, the
presence of the Higgs results in a small decrease in the total top-quark
cross section.

We also consider the effect of the pseudoscalar Higgs boson, $A$, on the
$t\bar t$ invariant-mass spectrum.  This particle does not couple to the
weak vector bosons, and couples to the top quark with standard-model
strength if the ratio of the vacuum-expectation values of the two Higgs
doublets is close to unity. For larger values of this ratio, the
pseudoscalar Higgs coupling to top quarks is suppressed, so its width is
narrower. The production of the pseudoscalar Higgs
via a bottom-quark loop and the decay to $b\bar b$ may also become significant,
reducing the branching
ratio to $t\bar t$ and the effect of the Higgs on top-quark production.
However,
the pseudoscalar Higgs does not couple to weak
vector bosons in a generic two-Higgs-doublet model, so there is no
competition from the decay to weak vector bosons.

The differential cross section for
$gg\rightarrow t\bar t$,
including the squares of the pseudoscalar-Higgs amplitude and the
continuum QCD amplitude, as well as the interference of the two amplitudes, is
\begin{eqnarray}
\frac{d\sigma}{dz} & = & \frac{3\alpha_s^2 G_F^2m^2s^2}{2048\pi^3}\beta
\left|\frac{P(s/m^2)}{s-m_A^2+im_A\Gamma_A(s)}\right|^2 \nonumber \\
& - & \frac{\alpha_s^2 G_F m^2s}{256\pi\sqrt 2}\beta \left(\frac{1}{p_1\cdot
p_3} + \frac{1}{p_2\cdot p_3}\right)
{\rm Re}\left[\frac{P(s/m^2)}{s-m_A^2+im_A\Gamma_A(s)}\right] \nonumber \\
& + & \frac{d\sigma_{QCD}}{dz}\;.
\end{eqnarray}
There is no interference between the scalar-Higgs and the
pseudoscalar-Higgs amplitudes.
The function associated with the virtual top-quark loop is
\begin{equation}
P(s/m^2)=-\;\frac{m^2}{s}I(s/m^2)
\end{equation}
where $I(s/m^2)$ is given above,
and the energy-dependent Higgs width is
\begin{equation}
m_A\Gamma_A(s)=\frac{3G_Fm^2s}{4\pi\sqrt 2}\beta\;.
\end{equation}
The resulting cross sections, given in Fig.~3, are qualitatively similar to the
cross sections in Fig.~2 for the scalar Higgs.  The pseudoscalar-Higgs width
is suppressed by $\beta$, rather than $\beta^3$ as is the scalar-Higgs width,
so the pseudoscalar Higgs boson is noticeably wider than the scalar Higgs boson
for $m_{H,A}=$ 400 and 500 GeV.

The peak-dip structure in the $t\bar t$ invariant-mass spectrum due to the
scalar or pseudoscalar Higgs bosons can be viewed as due to a final-state
interaction of the $t\bar t$ pair \cite{FSI,BB,MP,BBDKW}.  If one cuts
vertically through the top-quark loop in
Fig.~1(a), putting the cut propagators on shell, the amplitude factorizes into
the product of the amplitudes for the QCD
production of $t\bar t$ (Fig.~1(b)) and the
elastic scattering of the $t\bar t$ pair via the Higgs resonance, times a
phase-space factor $\beta$.  The
elastic $t\bar t$ partial-wave scattering amplitude may be represented by
a unitary phase-shift expression,
\begin{equation}
\frac{1}{\beta}\frac{-m_H\Gamma_H(s)}{s-m_H^2+im_H\Gamma_H(s)}
=\frac{1}{\beta}e^{i\delta}\sin\delta
\end{equation}
where $\delta(s)$ is the phase shift, which passes through $\pi/2$ at
$s=m_H^2$.  Using the Cutkosky rules, the absorptive part of the color-singlet,
zeroth-partial-wave Higgs amplitude is thus $a_0^{abs}=ia_0^{QCD}
e^{i\delta}\sin\delta$, where $a_0^{QCD}$ is the color-singlet,
zeroth partial wave of the QCD amplitude.  The total color-singlet,
zeroth-paritial-wave amplitude is the sum of the QCD amplitude and the
Higgs amplitude,
\begin{eqnarray}
a_0 & = & a_0^{QCD}+ia_0^{QCD}e^{i\delta}\sin\delta
+a_0^{disp}e^{i\delta}\sin\delta \nonumber \\
& = & e^{i\delta}[a_0^{QCD}\cos\delta + a_0^{disp}\sin\delta] \nonumber \\
& = & a_0^{QCD}\frac{s-m_H^2}{s-m_H^2+im_H\Gamma_H(s)}
+a_0^{disp}\frac{-m_H\Gamma_H(s)}{s-m_H^2+im_H\Gamma_H(s)}
\end{eqnarray}
where $a_0^{disp}$ is the one-loop-induced coupling of the
Higgs to the initial gluons due to the dispersive part of the top-quark
loop.  We see that the interference of the absorptive part of the loop
diagram with the tree diagram yields an exact zero at the resonance energy,
$s=m_H^2$ ($\delta=\pi/2$).  The one-loop-induced coupling due to the
dispersive part of the loop produces
the usual resonant peak, so generically a peak-dip structure results.
If the dispersive part of the loop is small compared with the absorptive part,
the dip will be the dominant
feature of the structure in the invariant-mass spectrum.  This is what
occurs for $m_{H,A}=$ 700 and 800 GeV in Figs.~2 and 3.

Peak-dip structures due to final-state interactions are well-known in
hadronic physics \cite{FSI,BB,MP}.  An example of a dip is the effect of the
$f_0(975)$ on the $\pi\pi$ invariant-mass
spectrum in double-diffractive pion production \cite{AKESSON}.
The argument in the preceding paragraph shows that a peak-dip structure in the
$t\bar t$ invariant-mass spectrum results whenever there is a resonant
final-state interaction
of the $t\bar t$ pair, regardless of the physics which produces the
resonant final-state interaction.\footnote{However, a spin-one resonance
cannot effect the $t\bar t$ invariant-mass distribution at leading order,
because the incoming gluons are forbidden to form a state of unit total angular
momentum by Yang's theorem \cite{YANG}.}  It also shows that the peak-dip
structure is not washed out by QCD corrections, since it depends only on
the final-state interaction of the $t\bar t$ pair, whose production may in
principle be calculated to any order in QCD.

At the Fermilab Tevatron, a top quark of mass greater than 150 GeV is
predominantly produced via quark-antiquark annihilation.  For example,
for $m_t=$ 170 GeV, the gluon-fusion contribution to the total top-quark
cross section is about $20\%$ \cite{BTG}.  Thus any structure in the
$gg\rightarrow
t\bar t$ cross section due to a Higgs boson would be buried beneath the
$q\bar q \rightarrow t\bar t$ continuum, and require very large statistics
to uncover. In contrast, at a higher-energy hadron collider, such as the
CERN Large Hadron Collider (LHC),
the gluon-fusion process is the
dominant source of top quarks. At a 4 TeV $p\bar p$
collider, the gluon-fusion process accounts for
about half of the total top-quark cross section (for $m_t\approx $ 170 GeV).

The $t\bar t$ invariant mass can be reconstructed at hadron colliders via
$t\bar t\to b\bar b W^+W^-$, followed by hadronic decay of one $W$ boson
and leptonic decay of the other.  The presence  of two $W$ bosons
and two top quarks constrains the kinematics sufficiently that the event can be
fully reconstructed, in principle.  We assume in the following
that the $t\bar t$ invariant mass can be reconstructed with reasonable
resolution.

The effect of the scalar and pseudoscalar Higgs bosons on the $t\bar t$
invariant-mass distribution is quite small, and will require large
statistics to observe.  For example, the peak and dip in the invariant-mass
distribution in Fig.~2 due to a scalar Higgs of mass 500 GeV are each about
a $3.5\%$ effect, each contained in a bin of width about 25 GeV.
The statistical significance of the peak
and dip depend on the number of $t\bar t$ events in the mass bin containing
the signal.  We present in Fig.~4 the
luminosity function of gluon-gluon collisions at hadron colliders of
various energy.  Multiplication of the $gg\to t\bar t$ cross section in Figs.~2
and 3 at a given invariant mass, $\sqrt s=M$, by the luminosity function at the
same invariant mass yields the inclusive hadronic differential cross section,
$d\sigma/dM$, for $pp,p\bar p\to t\bar t+X$.  At the LHC, with 100 $fb^{-1}$ of
integrated luminosity, the raw number of events in an invariant-mass
bin of width 25 GeV at $M=$ 500 GeV is about $4\times 10^6$.  Including a
branching ratio for the one hadronic - one leptonic decay of the $W$ bosons
of about 0.3, and a factor of 1/3 to account for acceptance and efficiencies,
leaves approximately $4\times 10^5$ events.  This is far more than
necessary to detect a $3.5\%$ effect at the $5\sigma$ level.  For
a scalar Higgs of mass 800 GeV, the dip corresponds to a $1.5\%$ decrease
in the number of events in a bin of width 50 GeV.  The number of
events, including branching ratios, acceptance, and efficiencies, is about
$6\times 10^4$, yielding a significance of about 3.5$\sigma$ for the signal,
marginally significant.  The results for the pseudoscalar Higgs are
comparable.

Lower-energy machines also have the potential to observe a
signal, depending on the luminosity and the Higgs mass.  A 4 TeV $p\bar p$
collider requires about 100 $fb^{-1}$ of integrated luminosity to detect a
scalar Higgs of mass 500 GeV with a 5$\sigma$ significance.  However, the
Fermilab Tevatron would require an integrated luminosity far in excess of
100 $fb^{-1}$ to detect such a Higgs boson.

In contrast to the class of models being considered here,
in the standard Higgs model the branching ratio of the Higgs boson to
$t\bar t$ is much less than unity.  For $m_t=$ 170 GeV, the branching ratio
is at most $18\%$ (this value
occurs for $m_H\approx$ 500 GeV). The Higgs boson produces only a tiny
wiggle in the $t\bar t$ invariant-mass spectrum at the Higgs mass,
too small to appear in Figs.~2 or 3.

We conclude that the peak-dip structure in the $gg\to t\bar t$ invariant mass
distribution, due to a scalar or pseudoscalar Higgs boson decaying
principally to top quarks, is potentially observable at the LHC, if the
$t\bar t$ invariant-mass resolution is less than the width of the Higgs boson.
Although the standard-model Higgs boson is not of this type, Higgs
bosons which decay predominantly to top quarks are natural in
two-Higgs-doublet models, in particular the heavy scalar Higgs, $H$, and the
pseudoscalar Higgs, $A$, of the minimal supersymmetric model.  The peak-dip
structure is unfortunately not observable at the Fermilab Tevatron with 1
$fb^{-1}$ of integrated luminosity, due to
the limited statistics and the relatively large rate of production of top
quarks via quark-antiquark annihilation.

\section*{Acknowledgements}

We are grateful for conversations with M. Albrow and C. Chiu.
This work was supported in part by contracts DE-FG03-93ER40757 and
DE-AC02-76CH00016 with the U.S. Department of Energy.

\clearpage

\section*{Figure Captions}

\vrule height0pt
\vspace{-22pt}

\bigskip

\indent Fig.~1 - Feynman diagrams for $gg \rightarrow t\bar t$: (a)
$s$-channel Higgs scalar ($H$) or pseudoscalar ($A$) exchange via a top-quark
loop (crossed diagram not shown),
(b) leading-order QCD ($u$- and $s$-channel diagrams not shown).
\bigskip

Fig.~2 - Total cross section for $gg\rightarrow t\bar t$, for $m_t=$ 170
GeV, as a function of the $t\bar t$ invariant mass. The calculation
includes the effects of the heavy Higgs scalar ($H$) of the minimal
supersymmetric model (with the ratio of the vacuum-expectation values
of the two Higgs doublets close to unity),
as well as the continuum QCD production of $t\bar t$,
for $m_H=$ 400, 500, 600, 700, and 800 GeV.

Fig.~3 - Same as Fig.~2, but for the Higgs pseudoscalar ($A$).

Fig.~4 - Luminosity function for gluon-gluon collisions at hadron colliders
of various energy. Multiplication of the $gg\to t\bar t$ cross section in
Figs.~2
and 3 at a given invariant mass, $\sqrt s=M$, by the luminosity function at the
same invariant mass yields the inclusive hadronic differential cross section,
$d\sigma/dM$, for $pp,p\bar p\to t\bar t+X$. The gluon distribution
functions have been evaluated at $\mu=M$.

\vfill


\begin{thebibliography}{99}

\bibitem{BC} M.~Berger and M.~Chanowitz, Phys.\ Rev.\ Lett.\ {\bf68},
757 (1992).

\bibitem{HKS} H.~Haber, G.~Kane, and T.~Sterling, Nucl.\ Phys.\ {\bf
B161}, 493 (1979).

\bibitem{HHG} For a review, see J.~Gunion, H.~Haber, G.~Kane, and S.~Dawson,
{\it The Higgs Hunter's Guide} (Addison-Wesley, New York, 1990).

\bibitem{GGMN} H.~Georgi, S.~Glashow, M.~Machacek, and D.~Nanopoulos,
Phys.\ Rev.\ Lett.\ {\bf 40}, 692 (1978).

\bibitem{GH} K.~Gaemers and F.~Hoogeveen, Phys.\ Lett.\ {\bf 146B}, 347
(1984).

\bibitem{QCD} L.~Jones and H.~W.~Wyld, Phys.\ Rev.\ D {\bf 17}, 1782 (1978);
B.~Combridge, Nucl.\ Phys.\ {\bf B151}, 429 (1979);
R.~K.~Ellis and J.~Sexton, Nucl.\ Phys.\ {\bf B282}, 642 (1987).

\bibitem{EW} ALEPH, DELPHI, L3, OPAL, and the LEP Electroweak Working
Group, CERN-PPE-93-157 (1993).

\bibitem{BTG} F.~Berends, J.~Tausk, and W.~Giele, Phys.\ Rev.\ D {\bf 47},
2746 (1993).

\bibitem{FSI} L.~Resnick, Phys.\ Rev.\ D {\bf 2}, 1975 (1970);
 J.~Pumplin, Phys.\ Rev.\ D {\bf 2}, 1859 (1970);
T.~Bauer, Phys.\ Rev.\ Lett.\ {\bf 25}, 485 (1970).

\bibitem{BB} J.-L.~Basdevant and E.~Berger, Phys.\ Rev.\ D {\bf 16}, 657
(1977); D {\bf 19}, 239 (1979); D {\bf 19}, 246 (1979).

\bibitem{MP} D.~Morgan and M.~Pennington, Z.\ Phys.\ {\bf C37}, 431 (1988).

\bibitem{BBDKW} J-L.~Basdevant, E.~Berger, D.~Dicus, C.~Kao, and
S.~Willenbrock, Phys.\ Lett.\ {\bf 313B}, 402 (1993).

\bibitem{AKESSON} T.~{\AA}kesson {\it et al.}, Nucl.\ Phys.\ {\bf
B264}, 154 (1986).

\bibitem{YANG} C.~N.~Yang, Phys.\ Rev.\ {\bf 77}, 242 (1950).

\end{thebibliography}
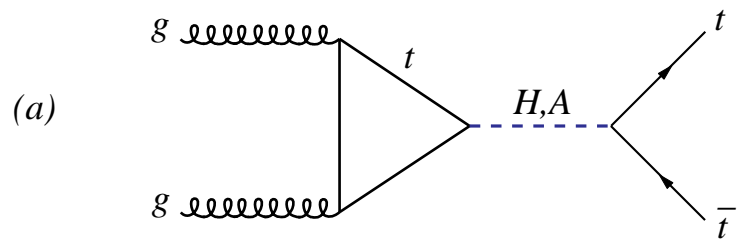
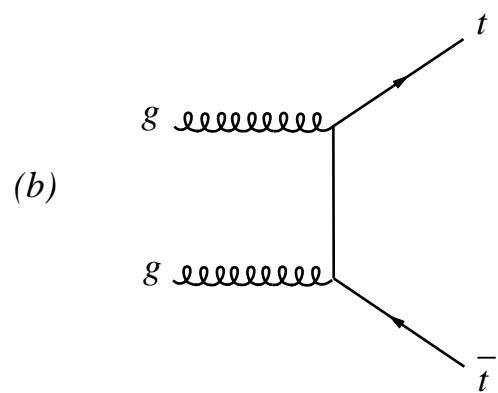
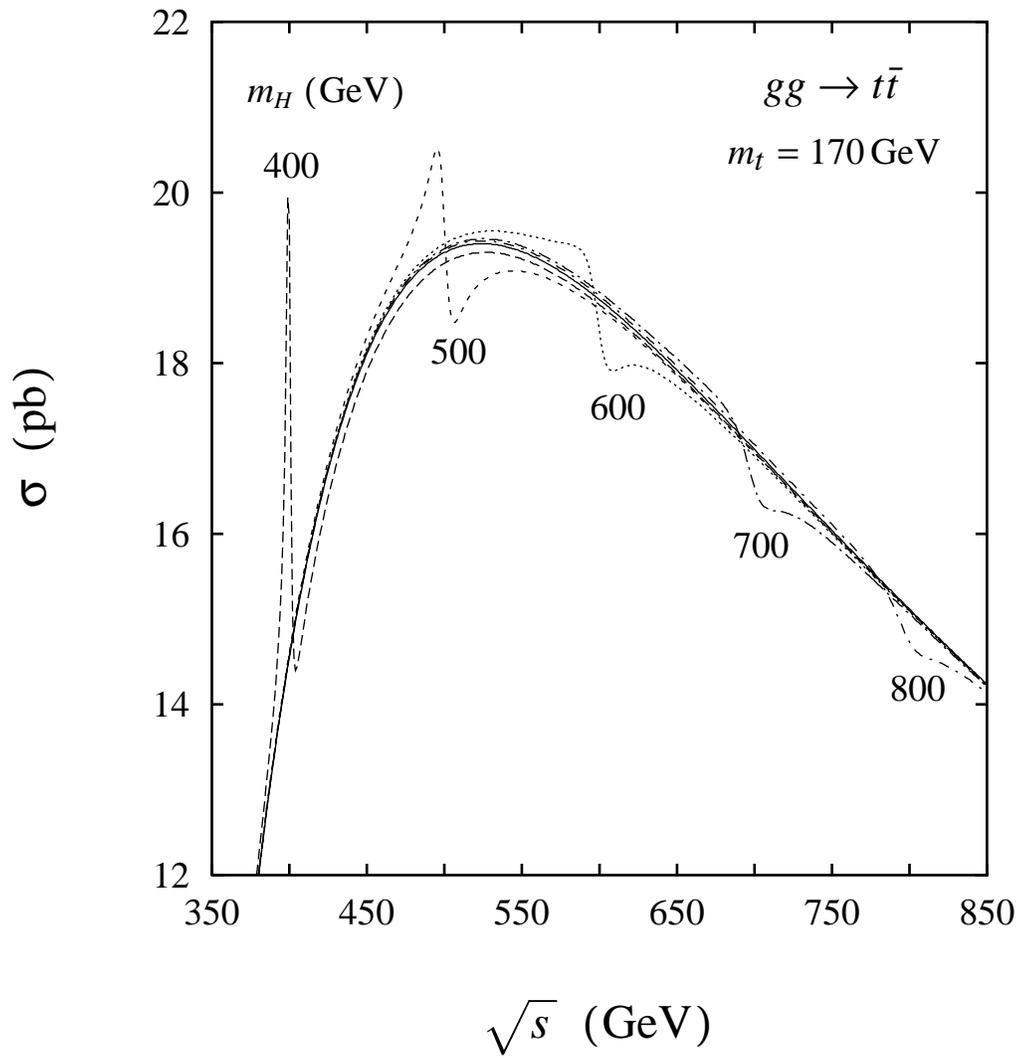
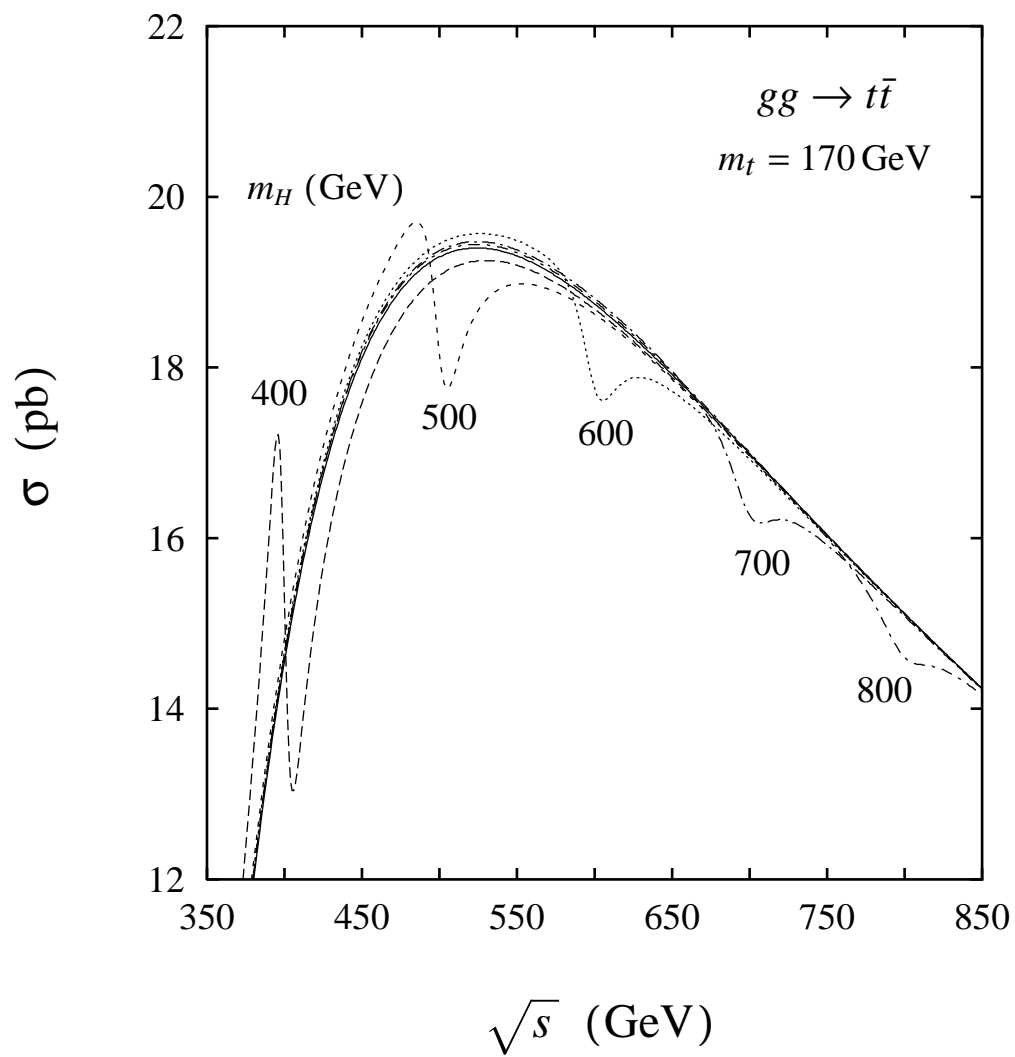
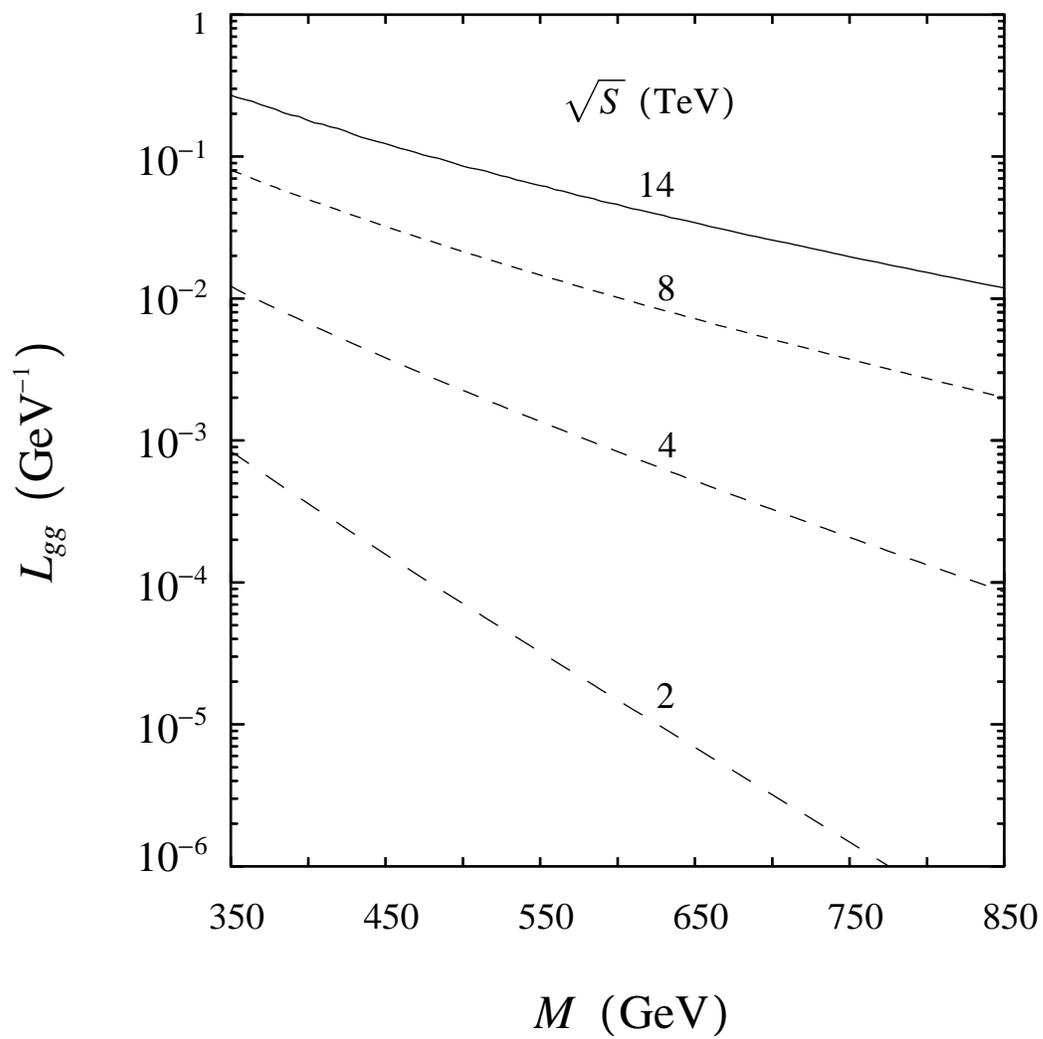
\end{document}